\documentclass[%
 reprint,
 amsmath,amssymb,
 aps,
]{revtex4-2}

\usepackage{graphicx}
\usepackage{dcolumn}
\usepackage{bm}

\usepackage{comment}  

\usepackage{xcolor} 
\usepackage{multirow}
\usepackage{xfrac}

\usepackage{ulem}

\usepackage{tikz, tkz-euclide, ifthen}
\definecolor{gold}{rgb}{0.95, 0.69, 0.24}
\definecolor{grey}{rgb}{0.57, 0.57, 0.57}

\pgfset{
	foreach/parallel foreach/.style args={#1in#2via#3}{evaluate=#3 as #1 using {{#2}[#3-1]}},
}

\definecolor{amethyst}{rgb}{0.6, 0.4, 0.8}

\begin{document}
\preprint{APS/123-QED}

\title{Plasma Frequency of Wire Medium Revisited}

\author{Denis Sakhno}
\email{denis.sakhno@metalab.ifmo.ru}

\author{Pavel A. Belov}

\affiliation{School of Physics and Engineering, ITMO University, Kronverksky Pr. 49, 197101, St. Petersburg, Russia}


\begin{abstract}

This paper revisits a model for the plasma frequency of a simple wire medium formed by a rectangular lattice of parallel metallic wires. We provide a comparative analysis of existing formulae for estimating the plasma frequency and derive a new expression. 
The proposed formula demonstrates superior accuracy for a square lattice of thin wires, with a relative error of less than $0.16\%$ for the ratio of wires radii to period smaller than $0.13$, significantly outperforming previously known results in this range. For rectangular lattices with the periods ratio from $2$ to $10$ the new formula provides estimations with a relative error less than $2.7\%$ for the ratio of wires radii to the smaller period less than $0.4$.

\end{abstract}

\maketitle



Wire metamaterials, also referred to as \textit{rodded media} (or rodded-type artificial dielectrics) in pioneering works \cite{brown1950design, brown1953artificial, carne1959theory, rotman1962plasma}, are artificially engineered structures composed of metallic wires or rods periodically arranged in space to form two- or three-dimensional lattices.
These materials have attracted significant attention since they feature plasma-like behavior \cite{rotman1962plasma, pendry1996extremely, pendry1998low, silveirinha2009artificial}, despite containing no free charges or actual plasma.
Wire metamaterials were initially proposed in the context of microwave lens design \cite{kock1946metal, kock1948metallic}, exhibiting a refractive index less than unity \cite{brown1953artificial, silveirinha2005homogenization}. They have also been extensively used as components in the design of left-handed media \cite{smith2000composite, shelby2001microwave, shelby2001experimental}, subwavelength imaging devices \cite{belov2006subwavelength, silveirinha2007imaging, shvets2007guiding, belov2010experimental} and to improve magnetic resonance imaging (MRI) performance \cite{slobozhanyuk2016enhancement, guo2024metamaterial}.

\begin{figure}[h!]
        \begin{minipage}{0.9\linewidth}
		\center{ 
			\usetikzlibrary{quotes,angles}

\pgfset{
	foreach/parallel foreach/.style args={#1in#2via#3}{evaluate=#3 as #1 using {{#2}[#3-1]}},
}
\definecolor{gold}{rgb}{0.95, 0.69, 0.24}
\definecolor{grey}{rgb}{0.57, 0.57, 0.57}

\begin{tikzpicture}[scale=0.98, transform shape]
	\def \c{1.75}  
	\def \a{2.3}  
	\def \ang{90}  
	
	\def \shiftx {1.5 * \a + 1.5 * \c * cos(\ang)}
	\def \shifty {0.5 * \c * sin(\ang)}
	
	\coordinate(A1) at ({0 + \shiftx}, {0 + \shifty}); 
	\coordinate(A2) at ({\a + \shiftx}, {0 + \shifty}); 
	\coordinate(A3) at ({\a + \c * cos(\ang) + \shiftx}, {\c * sin(\ang) + \shifty}); 
	\coordinate(A4) at ({\c * cos(\ang) + \shiftx}, {\c * sin(\ang) + \shifty}); 
	\def \verts {A1, A2, A3, A4}
	
	\draw[->, line width=0.5pt, grey] (0, 0) -- ({1 / 3 * \a}, 0) node[left, xshift=3pt, yshift=-7pt] {$x$};
	\draw[->, line width=0.5pt, grey] (0, 0) -- (0, {{\a * 1 / 3}}) node[left, , xshift=-1pt, yshift=-3pt] {$y$};
	
	\def \rw{0.05*\a}  
	\def \wirescol{gold}  
	
	\def \numa {3}  
	\def \numc {2}  
	\def \delta {0.3}  
	
	\foreach [evaluate={
		\linex = \c * cos(\ang) * \numc ;
		\liney = \c * sin(\ang) * \numc 
	}] \ia in {0, ..., \numa} {
	
		\coordinate(diag1) at ({ \ia * \a - \delta * cos(\ang) }, {- \delta * sin(\ang) });
		\coordinate(diag2) at ({ \ia * \a + \linex + \delta * cos(\ang) }, { \liney + \delta * sin(\ang) });
		
		\draw[dashed, line width=0.5pt, grey](diag1) -- (diag2);
	}
	
	\foreach [evaluate={
		\rowy = \c * sin(\ang) * \ic
	}] \ic in {0, ..., \numc} {
		
		\coordinate(left) at ({ -\delta }, { \rowy });
		\coordinate(right) at ({ (\numa ) * \a + \delta }, { \rowy });
		
		\draw[dashed, line width=0.5pt, grey](left) -- (right);
	}
	
	\draw[solid, line width=1.0pt, red](A1) -- (A2) -- (A3) -- (A4)--(A1);
        \def \vynos {0.25 * \a}
        \def \vynosCol {black}
        \def \vynosLines {solid}
        \draw[\vynosLines, line width=0.25pt, \vynosCol]({\shiftx}, {\shifty + \c}) -- ({\shiftx - \vynos}, {\shifty + \c});
        \draw[\vynosLines, line width=0.25pt, \vynosCol]({\shiftx}, {\shifty}) -- ({\shiftx - \vynos}, {\shifty});
        \def \vynos {0.25 * \a}
        \draw[\vynosLines, line width=0.25pt, \vynosCol]({\shiftx}, {\shifty}) -- ({\shiftx}, {\shifty - \vynos});
        \draw[\vynosLines, line width=0.25pt, \vynosCol]({\shiftx + \a}, {\shifty}) -- ({\shiftx + \a}, {\shifty - \vynos});

        \foreach \vert in \verts {
		\filldraw [red] (\vert) circle (0.5pt); 
	}
    
	
	\def \ic {0}  
	\def \numa {3}  
	\foreach [evaluate={
		\thisx = \a * \ia + \c * cos(\ang) * \ic ;
		\thisy = \c * sin(\ang) * \ic 
	}] \ia in {0, ..., \numa} {
			\coordinate(this) at ({ \thisx }, { \thisy });
			\filldraw [black, fill=\wirescol, line width=0.25pt] (this){} circle (\rw);
	}
	
	\def \ic {1}  
	\def \numa {3}   
	\foreach [evaluate={
		\thisx = \a * \ia + \c * cos(\ang) * \ic ;
		\thisy = \c * sin(\ang) * \ic 
	}] \ia in {0, ..., \numa} {
		\coordinate(this) at ({ \thisx }, { \thisy });
		\filldraw [black, fill=\wirescol, line width=0.25pt] (this){} circle (\rw);
	}
	
	\def \ic {2}  
	\def \numa {3}    
	\foreach [evaluate={
		\thisx = \a * \ia + \c * cos(\ang) * \ic ;
		\thisy = \c * sin(\ang) * \ic 
	}] \ia in {0, ..., \numa} {
		\coordinate(this) at ({ \thisx }, { \thisy });
		\filldraw [black, fill=\wirescol, line width=0.25pt] (this){} circle (\rw);
	}
	
	
	\def \circX {2 * \a + 1 * \c * cos(\ang)}
	\def \circY {1 * \c * sin(\ang)}
	\def \angArr {\ang + 45}

	\draw[->, black, line width=0.5pt]({\circX-3*\rw *cos(\angArr)}, {\circY-3*\rw * sin(\angArr)}) -- ({\circX-\rw*cos(\angArr)}, {\circY-\rw*sin(\angArr)}) node[midway,xshift=12pt, yshift=-3pt] {\large $2r_0$};
	\draw[->, black, line width=0.5pt]({\circX+3*\rw *cos(\angArr)}, {\circY+3*\rw * sin(\angArr)}) -- ({\circX+\rw*cos(\angArr)}, {\circY+\rw*sin(\angArr)});
	
        \def \xArrowLeftX {1.5}
        \def \xArrowY {0.35}
	\coordinate (aLeft) at ({\xArrowLeftX * \a}, {\xArrowY * \c * sin(\ang});
	\coordinate (aRight) at ({\a + \xArrowLeftX * \a}, {\xArrowY * \c * sin(\ang});
	\draw[<->, line width=0.5pt, black] (aLeft) -- (aRight) node[midway, below, sloped, xshift=10pt, yshift=0pt] {\large $a$};

        \def \thetaArrowLeftX {1.35}
        \def \thetaArrowY {0.5}
	\coordinate (cLow) at ({\thetaArrowLeftX * \a}, {\thetaArrowY * \c * sin(\ang});
	\coordinate (cUp) at ({\thetaArrowLeftX * \a}, {(\thetaArrowY + 1) * \c * sin(\ang});
	\draw[<->, line width=0.5pt, black] (cLow) -- (cUp) node[left, sloped, xshift=0pt, yshift=-15pt] {\large $b$};
	
\end{tikzpicture}
		}
	\end{minipage}
	\caption{
		Geometry of a simple wire metamaterial formed by a rectangular lattice (with periods $a$ and $b$) of parallel wires of the radii equal to $r_0$. A unit cell is highlighted.
	} \label{fig:square_swm} 
\end{figure}

Recently, interest in wire media has been renewed due to its application in dark matter searches \cite{millar2023alpha}, where a simple wire medium was used as the main component of a plasma haloscope. The plasma frequency, which defines the resonant frequency of the proposed haloscope design, becomes a crucial parameter requiring precise estimation. For this reason, in this paper, we revisit analytical formulae for the plasma frequency estimation, specifically for a simple wire metamaterial formed by a rectangular lattice of identical wires in free space (see Fig. \ref{fig:square_swm}).

Homogenization of a wire medium (Fig. \ref{fig:square_swm}) composed of dispersive rods ($\varepsilon_\text{rods}$) was performed in \cite{silveirinha2006homogenization} resulting in an effective permittivity tensor
\begin{equation}
    \overline { \overline{\varepsilon} } = \operatorname{diag} \left( 1, 1, \varepsilon_{zz} (\omega, q_z) \right),
\end{equation}
\begin{equation}
    \varepsilon_{zz} =
    1 + \left(
    \frac{1}{\varepsilon_\text{rods}(\omega)-1} \frac{ab}{\pi r_0^2} - 
    \frac{ \left(\frac{\omega}{c}\right)^2 - q_z^2}{k_p^2(a,b,r_0)}
    \right)^{-1},
    \label{eq:eps_zz_0}
\end{equation}
where $c$ is the speed of light in a vacuum, $\omega$ is a wave frequency, $q_z$ is a wavevector projection on the wires direction and $k_p=\omega_p/c$ is the plasma wavenumber ($\omega_p$ is the plasma frequency), which depends only on geometric parameters of a wire medium: lattice periods $a$ and $b$, and radius of wires $r_0$.

For plasmonic rods with $\varepsilon_\text{rods}=1-\omega_{p,\text{ rods}}^2/\omega^2$ Eq. (\ref{eq:eps_zz_0}) in $\Gamma$-point reduces to Drude formula with plasma frequency $\omega_{p,\text{ eff}}$ given by following expression
\begin{equation}
    \frac{1}{\omega_{p, \text{ eff}}^2} = \frac{1}{\omega_p^2} + \frac{1}{\omega_{p,\text{rods}}^2} \frac{ab}{\pi r_0^2}.
\end{equation}

At microwave frequency range metal wires can be well approximated as perfectly conducting (PEC, $\varepsilon_\text{rods}\rightarrow -\infty$) for which Eq. (\ref{eq:eps_zz_0}) reduces to
\begin{equation}
    \varepsilon_{zz}^{(\text{PEC})} =
    1 - \frac{k_p^2(a,b,r_0)}{ \left(\frac{\omega}{c}\right)^2 - q_z^2},
    \label{eq:eps_zz_1}
\end{equation}
which is equal to $0$ at the plasma frequency $\omega_p$ in $\Gamma$-point ($q_z=0$).





The plasma frequency of a wire medium composed of PEC wires can be also defined as a cut-off frequency for electromagnetic waves whose electric field is aligned with the wires (TM-polarized).
In Table \ref{tab:overview}, various equations and estimation formulae (approximations) for the plasma frequency of rectangular and square lattices of parallel metallic wires (rewritten using our notation proposed in Fig. \ref{fig:square_swm}) are presented.

\begin{table*}[t]
  \centering
  \begin{tabular}{|l|c||cc|}
    \hline \rule{0pt}{10pt}
    Authors & Ref. & 
    Equations and Estimations for $k_p^2$ &  \\
    \hline
    \hline \rule{0pt}{20pt}
    J. Brown (equation) & \cite{brown1953artificial} & 
    $\displaystyle \left(k_p b\right) \; \tan \left( \frac{k_p b}{2} \right) = \frac{b}{a} \frac{\pi}{\ln \frac{a}{2\pi r_0}} $ 
    & (I.1) \\ \rule{0pt}{30pt}
    P.A. Belov (equation) & \cite{belov2002dispersion} & 
    $ \displaystyle
    \frac{1}{\pi}\ln{\frac{a}{2\pi r_0}} -
    \frac{1}{k_p a} \cot \left( \frac{k_p b}{2} \right)
    + 
    \sum\limits_{n=1}^{\infty}
    \frac{1}{\pi n} \left(
    \frac{
    \coth \left( \pi n \frac{b}{a} \; \sqrt{1-\left( k_p a / 2\pi n \right)^2} \right) 
    }{ 
    \sqrt{1-\left( k_p a / 2\pi n \right)^2} 
    } - 1 \right)=0 $
    & (I.2) \\
    & & & \\ \hline \hline \rule{0pt}{20pt}
    J.B. Pendry et al. & \cite{pendry1996extremely, pendry1998low} & 
    $\displaystyle k_p^2 =\frac{ 2\pi / a^2 }{ \ln \frac{a}{r_0} } $ 
    & (I.3) \\ \rule{0pt}{20pt}
    A.K. Sarychev et al. & \cite{sarychev2001commentpaperextremelylow} &
    $ \displaystyle k_p^2 = \frac{2\pi / a^2}{ \ln \frac{a}{\sqrt{2} r_0} + \frac{\pi}{4} -\frac{3}{2} }$ 
    & (I.4) \\ \rule{0pt}{20pt}
    P.A. Belov et al. & \cite{belov2002dispersion, belov2003strong} & 
    $\displaystyle k_p^2 = \frac{2\pi/a^2}{ \ln \frac{a}{2\pi r_0} + \frac{\pi}{6} + 
    \sum\limits_{n=1}^{\infty} \frac{\coth (\pi n) - 1}{n}} \approx
    \left. \frac{2\pi/a^2}{\ln \frac{a}{2\pi r_0} + 0.5275} \right.^* $
    & (I.5) \\ \rule{0pt}{20pt}
    G.B. Shvets et al. & \cite{shvets2003electromagnetic} & 
    $\displaystyle k_p^2 = \frac{ 8/a^2 }{ \ln \frac{a}{2\sqrt{2} r_0} } $ 
    & (I.6) \\ \rule{0pt}{20pt}
    A.V. Tyukhtin et al. & \cite{tyukhtin2011effective} & 
    $\displaystyle k_p^2 = \frac{2\pi /a}{ \ln \frac{a}{r_0} - 1.0487} $
    & (I.7) \\ 
    & & & \\ \hline \hline \rule{0pt}{20pt}
    S.I. Maslovski et al. & \cite{maslovski2002wire, maslovski2009nonlocal} & 
    $\displaystyle k_p^2 = \frac{2\pi / a^2}{ \ln \frac{a^2}{4r_0(a-r_0)} } $ 
    & (I.8) \\ \rule{0pt}{20pt}
    A. Kumar et al. & \cite{kumar2012novel} & 
    $\displaystyle k_p^2 = \frac{2\pi}{a^2} \times \left\{ 
    1.763 \frac{r_0}{2a} +
    \left[ 
    1.264 + 
    \ln \frac{a^2}{4r_0 \left( \sqrt{2} a - r_0 \right)}
    \right] - 
    \frac{d}{a}
    \left[
    \arctan \frac{r_0}{\sqrt{2} d}
    +
    \arctan \frac{a}{d}
    \right]
    \right\}^{-1}
    $ 
    & (I.9) \\ \rule{0pt}{15pt}
    & & \multicolumn{1}{l}{ where $d=\sqrt{a^2 - r_0^2}$ } & \\
    & & & \\
    \hline
  \end{tabular}
  \caption{
  Analytical formulae for the plasma wavenumber $k_p$ of a wire medium, including references to the earlier works in which they were derived. 
  Formulae (I.1–2) correspond to a wire medium with \textbf{a rectangular lattice}, whereas formulae (I.3–9) correspond to one with \textbf{a square lattice}.
  \\
  $^*$ Note that in \cite{belov2002dispersion} and in most works referring to it the value of the denominator constant (I.5) was slightly misestimated as $0.5275$, since the more accurate value is $0.5273$. Nevertheless, it had little impact on the estimation accuracy of formula (I.5).
  }
  \label{tab:overview}
\end{table*}

Equation (I.1), introduced by J. Brown in 1953 \cite{brown1953artificial} appears to be the earliest known result for the cut-off frequency of a system of "perfectly conducting cylinders arranged in a rectangular lattice". However, this formulation has not gained widespread use for estimating the plasma frequency, as it is not a direct formula for $k_p$, but rather a transcendental equation. 
 
In 1996-1998, interest in "thin-wire structures" as artificial plasma was revived by J.B. Pendry et al. \cite{pendry1996extremely, pendry1998low} and Eq. (I.3) for the plasma frequency was presented. Other formulations (I.2, 4-9), listed in Table \ref{tab:overview}, were proposed later (2001-2012) as part of a renewed focus on metamaterials and periodic wire structures in particular. The diversity among these estimations has led to some disagreement regarding which formula is most appropriate for experimental use. However, only a few articles \cite{brand2007complex, kumar2012novel} have attempted to systematically compare some of these formulae for a wire medium with a square unit cell.

\begin{figure*}[t]
        \begin{minipage}{0.485\linewidth}
		\center{ 
                \includegraphics[width=1.00\textwidth]{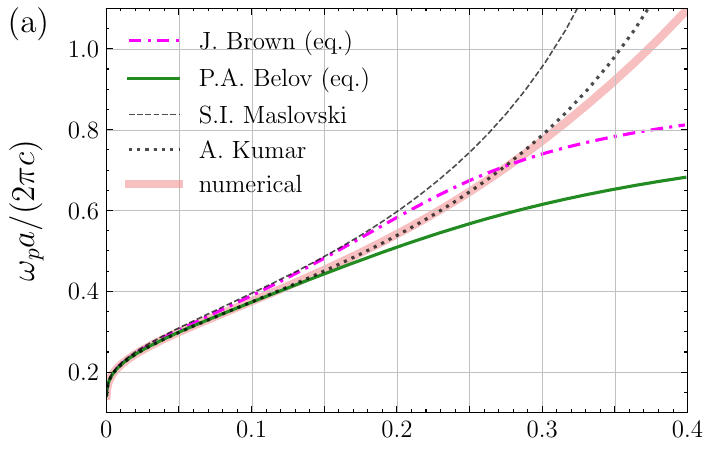}
		}
	\end{minipage}
        \hfill
        \begin{minipage}{0.485\linewidth}
		\center{ 
			\includegraphics[width=1.00\textwidth]{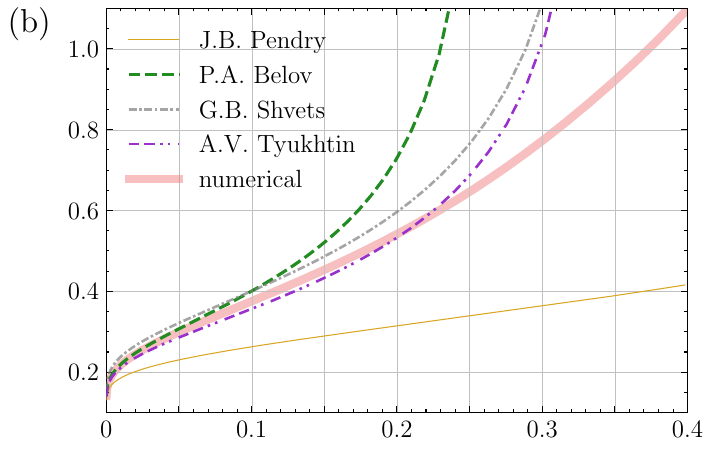}
		}
	\end{minipage}
        \vfill
        \begin{minipage}{0.485\linewidth}
		\center{ 
                \includegraphics[width=1.00\textwidth]{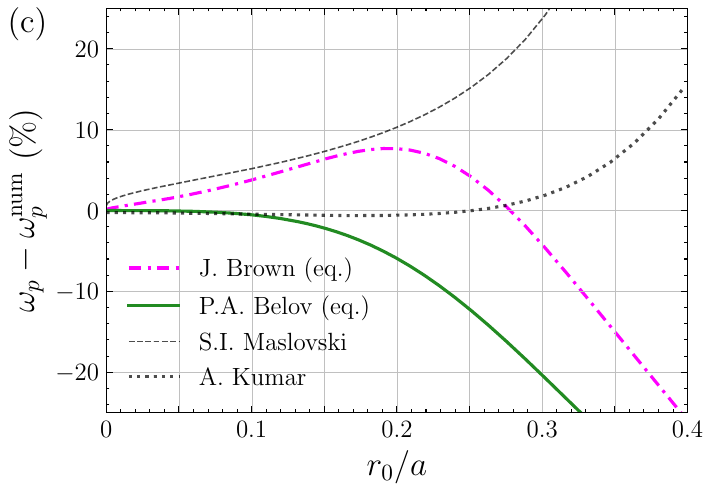}
		}
	\end{minipage}
        \hfill
        \begin{minipage}{0.485\linewidth}
		\center{ 
			\includegraphics[width=1.00\textwidth]{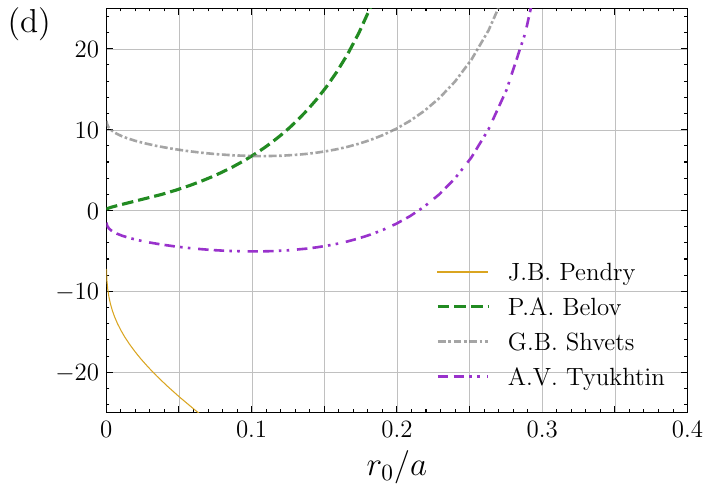}
		}
	\end{minipage}
	\caption{
		(a–b) Plasma frequency of a wire medium with \textbf{a square lattice} as a function of the $r_0/a$ ratio. The solid thick line corresponds to \textit{the exact value} (obtained numerically), while the other lines represent different estimation formulae from Table \ref{tab:overview}.
        (c–d) Relative error (in percent) with respect to \textit{the exact results} for the various formulae used to estimate the plasma frequency.
	} \label{fig:omega_comsol} 
\end{figure*}

First, we compare formulae performances for \textbf{a square lattice} of wires, since the most results given in Table \ref{tab:overview} in a closed form was derived for this geometry. For our comparison, we performed numerical simulations in COMSOL Multiphysics \cite{comsol} to calculate the plasma frequency for parallel PEC wires arranged in a square lattice ($a=b$, Fig. \ref{fig:square_swm}) varying the wire radii. The first eigenmode at the $\Gamma$-point was determined using an eigenmode solver applied to a unit cell with periodic boundary conditions.  These simulation results are used as \textit{the exact values} of the plasma frequency (the data file is provided in the Supplementary Material).
It is worth noting that recent studies on plasma haloscopes typically involve periods on the order of a few centimeters, while wire radii range from about $50$ $\mu$m \cite{kowitt2023tunable} to several millimeters \cite{enriquez2025uniform, sakhno2025honeycomb}.

In Figures \ref{fig:omega_comsol}(a) and (b), the thick solid line represents the numerically obtained plasma frequency function (\textit{the exact function}) of the $r_0/a$ ratio. 
The other curves correspond to analytical results derived using the formulae and equations listed in Table \ref{tab:overview}. We did not include all estimations from the table in a single plot to avoid overlapping curves and to enhance readability.


In Figure \ref{fig:omega_comsol}(a), we compare several formulae from Table \ref{tab:overview}, including the solutions of two transcendental equations (I.1-2).
To provide a clearer comparison of the formulae's performance, we plot the relative difference (in percent) between the analytical and \textit{exact} results as a function of the $r_0/a$ ratio in Fig. \ref{fig:omega_comsol}(c). 
This plot shows a great performance of the transcendental equation (I.2) by P.A. Belov et al., with an error of less than $0.5\%$ for thin wires ($r_0/a < 0.1$), and high accuracy of formula (I.9) by A. Kumar et al., maintaining an error below $2.5\%$ up to $r_0/a=0.3$.
Notably, Equation (I.1) by J. Brown performs better than most of the more recent estimations of the plasma frequency for square-lattice-based wire (rodded) media listed in Table \ref{tab:overview}, yielding a relative error of less than $8\%$ up to an $r_0$ value of approximately $0.32a$.

For many applications, such as dark matter searches, the accuracy of plasma frequency estimation must be better than $0.5\%$. This is the reason why the Eq. (I.2) by P.A. Belov et al. can be used only for $r_0/a < 0.1$. Meanwhile, the estimation given by Eq. (I.9) from A. Kumar et al. yields a nearly constant deviation of approximately $0.5\%$ for all radii smaller than $0.1a$.



\begin{table}[h!]
  \centering
  \begin{tabular}{|l|c||c|r|}
    \hline \rule{0pt}{12pt}
    Authors & Ref. & 
    $A$ & $C$\\
    \hline
    \hline \rule{0pt}{12pt}
    J.B. Pendry et al. & \cite{pendry1996extremely, pendry1998low} & 
    $ 2\pi / a^2 $ & $0$ \\ \rule{0pt}{12pt}
    A.K. Sarychev et al. & \cite{sarychev2001commentpaperextremelylow} &
    $ 2\pi / a^2 $ & $-1.0612$ \\ \rule{0pt}{12pt}
    P.A. Belov et al. & \cite{belov2002dispersion} & 
    $ 2\pi / a^2 $ & $^*-1.3105$ \\ \rule{0pt}{12pt}
    G.B. Shvets et al. & \cite{shvets2003electromagnetic} & 
    $ 8 / a^2 $ & $-1.0397$ \\ \rule{0pt}{12pt}
    A.V. Tyukhtin et al. & \cite{tyukhtin2011effective} & 
    $ 2\pi / a^2 $ & $-1.0487$ \\ 
    \hline
  \end{tabular}
  \caption{
  Table of parammeters $A$ and $C$ values for different formulae for the plasma frequency obtained in earlier works which can be written in the form of Eq. (\ref{eq:square_const}). \\
  $^*$ The constant value in the estimation by P.A. Belov et al. (I.5) was corrected here.}
  \label{tab:constants}
\end{table}

In Fig. \ref{fig:omega_comsol}(b) we have plotted analytical curves given by
formulae (I.3-7). All these formulae can be expressed in the following common form:
\begin{equation}
    k_p^2=\frac{A}{\ln \frac{a}{r_0} + C}.
    \label{eq:square_const}
\end{equation}
The corresponding values of $A$ and $C$ for these formulae, rewritten in this form, are provided in Table \ref{tab:constants}. Note that the values of $A$ and $C$ for the formulae by A.K. Sarychev (I.4) and A.V. Tyukhtin (I.7) are very similar; therefore, we include only one of them in the plot.

All estimations (I.3–I.7) were derived under the assumption of \textit{thin wires}. As a result, these formulae exhibit significant deviation from \textit{the exact values} for \textit{thick wires} ($a/r_0 > 0.1$), as shown in Fig. \ref{fig:omega_comsol}(d). For each formula, there is a rapid increase in error, reaching values on the order of several percent for thick wires.

\begin{figure}[h!]
    \begin{minipage}{0.97\linewidth}
        \center{
            \includegraphics[width=1.00\textwidth]{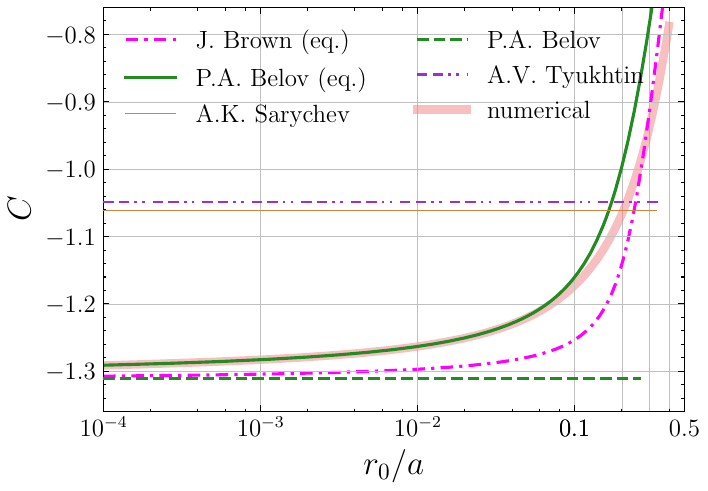}
        }
    \end{minipage}
    \caption{
    Dependence of the parameter $C$ from Eq. (\ref{eq:square_const}) on the wire radius $r_0$, extracted from numerical results (\textit{the exact results}) using Eq. (\ref{eq:extract_c}) -- thick solid line. Other curves represent parameter $C$ dependencies extracted from different formulae (see Table \ref{tab:overview}).
    } \label{fig:constants_compare} 
\end{figure}

In order to determine the cause of this deviation, we inverted Eq. (\ref{eq:square_const}) and calculated $C$ (using $A = 2\pi/a^2$, the most common value in Table \ref{tab:constants}) for all considered formulae:
\begin{equation}
    C(a, r_0)=\frac{2\pi}{(k_pa)^2} - \ln \frac{a}{r_0}.
    \label{eq:extract_c}
\end{equation}
Some of the extracted values of $C$ are plotted in Fig. \ref{fig:constants_compare}: (1) several horizontal lines corresponding to expressions (I.4), (I.5), and (I.7), which follow the form of Eq. (\ref{eq:square_const}); and (2) non-horizontal curves for the transcendental equations (I.1–I.2), obtained by applying the inverse function (\ref{eq:extract_c}) to their numerical solutions.

One can observe that $C$ corresponding to \textit{the exact results} (thick solid line) varies from $-1.29$ to $-0.78$. This indicates that treating $C$ as a constant cannot accurately describe the plasma frequency across a wide range of radii.
The value of $C$ given by Eq. (I.5) slightly underestimates the actual value (by about $0.02$) for extremely small radii, but its deviation becomes significant for larger radii. Meanwhile, the value of $C$ extracted from the transcendental equation (I.2) closely follows \textit{the exact curve} up to $r_0/a = 0.1$.

Since Eq. (I.2) is not straightforward to use in practical calculations, we propose deriving a simplified, yet accurate, approximation suitable for practical applications.






\begin{figure*}[t]
    \begin{minipage}{0.49\linewidth}
        \center{
            \includegraphics[width=1.00\textwidth]{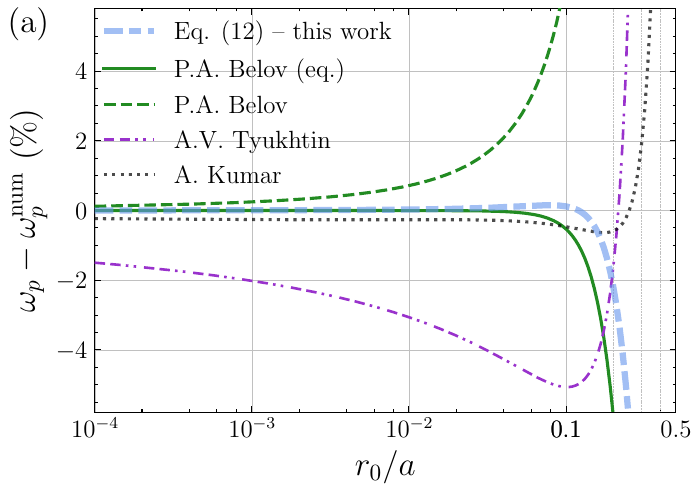}
        }
    \end{minipage}
    \hfill
    \begin{minipage}{0.499\linewidth}
        \center{
            \includegraphics[width=1.00\textwidth]{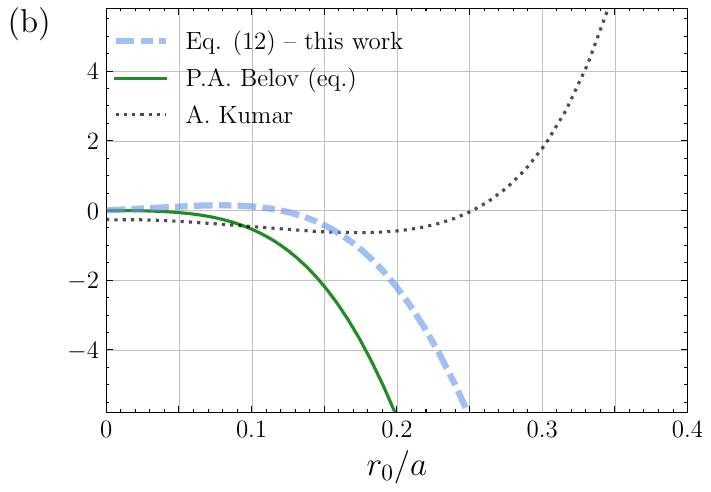}
        }
    \end{minipage}
    \caption{
    Comparison of different formulae from Table \ref{tab:overview} and the obtained formula (\ref{eq:square_new}) with \textit{the exact} plasma frequency \textit{values} for a wire medium with \textbf{a square lattice}, presented as relative errors plotted against wire radii for two scales: (a) thin wires and (b) thick wires.
    } \label{fig:eps_compare} 
\end{figure*}

\begin{figure}[h!]
    \begin{minipage}{0.81\linewidth}
        \center{
            \includegraphics[width=1.00\textwidth]{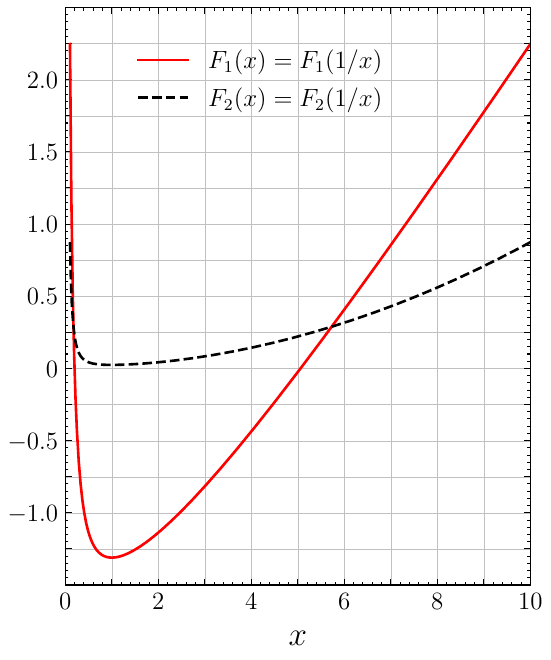}
        }
    \end{minipage}
    \caption{
    Functions $F_1(x)$ and $F_2(x)$ defined in Eqs. (\ref{eq:f1}-\ref{eq:f2}).
    } \label{fig:func} 
\end{figure}

A new formula for the plasma frequency can be derived 
for a rectangular lattice of wires ($a\times b$) via expanding Eq. (I.2) in a Taylor series up to the second power of $k_p a$ and $k_p b$ (since for thicker wires the value of $k_p$ is not as small as needed for the first order approximation \cite{sakhno2024anisotropy}):
\begin{align}
    \frac{1}{\pi}\ln{\frac{a}{2\pi r_0}} -
    & \frac{1}{k_p a} \left( \frac{2}{k_pb} - \frac{k_pb}{6} - \frac{\left(k_pb\right)^3}{360} \right)
    +
    \label{eq:square_disp_gamma_2}
    \\
    &\sum\limits_{n=1}^\infty  \frac{\coth{ \left( \pi n \frac{b}{a} \right) - 1} }{\pi n}
    + \nonumber \\
    \frac{(k_pa)^2}{8}
    &\sum\limits_{n=1}^\infty \left( \frac{\sinh^{-2} \left(\pi n \frac{b}{a} \right)}{(\pi n)^2} \frac{b}{a} + \frac{\coth\left(\pi n\frac{b}{a} \right)}{(\pi n)^3} \right)
    =0.
    \nonumber
\end{align}
Combining all same powers of $k_p \sqrt{ab}$ together after a multiplication of Eq. (\ref{eq:square_disp_gamma_2}) by $\pi (k_p \sqrt{ab})^2$ we obtain:
\begin{align}
    & (k_p^2ab)^2 \; F_2 (b/a) + \label{eq:square_disp_gamma_3} \\
    &(k_p^2ab)^1 \; 
    \Bigg[\ln{\frac{\sqrt{ab}}{r_0}} + F_1(b/a) \Bigg] - 
    2\pi
    =0,
    \nonumber
\end{align}
where 
\begin{align}
    F_1\left(x\right)=
    -\frac{1}{2}\ln x - \ln (2\pi) +&
    \frac{\pi x}{6}+ 
    \label{eq:f1} \\
    &\sum\limits_{n=1}^\infty  \frac{\coth{ \left( \pi n x \right) - 1} }{n}
    \nonumber
\end{align}
and
\begin{align}
    F_2\left(x\right)=&\frac{\pi}{8}\frac{1}{x} \Bigg[
    \frac{x^3}{45} +
    \label{eq:f2} \\
    &
    \sum\limits_{n=1}^\infty \left( \frac{\sinh^{-2} \left(\pi n x \right)}{(\pi n)^2} x + \frac{\coth\left(\pi n x \right)}{(\pi n)^3} \right)
    \Bigg]
    \nonumber
\end{align}
are the functions that are satisfy the equality $F_m(x)=F_m(1/x)$ ($m=1,2$).
These functions are plotted in Fig. \ref{fig:func}.
The only difference of Eq. (\ref{eq:square_disp_gamma_3}) with Eqs. (37-38) from \cite{belov2002dispersion} is in the first term of the fourth order (omitting the fourth-order term yields the same result as reported in \cite{belov2002dispersion}). Hence, the new formula for the plasma frequency estimation can be written as the positive solution of the quadratic equation (\ref{eq:square_disp_gamma_3}):
\begin{align}
    k_p^2=
    \frac{1}{2ab\; F_2(b/a)} 
    & \Bigg\{
    -
    \left[\ln \frac{\sqrt{ab}}{r_0} + F_1(b/a) \right]+ \label{eq:rect_root} 
    \\
    &
    \sqrt{
    \left[\ln \frac{\sqrt{ab}}{r_0} + F_1(b/a) \right]^2 +8\pi F_2(b/a)
    }
    \Bigg\}. \nonumber
\end{align}

For \textbf{a square wire medium} ($a=b$), after a substitution of numerical values of the functions introduced above ($F_1(1)\approx -1.3105$, $F_2(1)\approx 2.42967\cdot 10^{-2}$), Eq. (\ref{eq:rect_root}) can be written in \textit{a ready-to-use form}:
\begin{align}
    k_p^2 = & \frac{1}{4.8593 \cdot 10^{-2}\; a^2} \;
    \Bigg\{ -\left[\ln \frac{a}{r_0} - 1.3105\right] + 
    \label{eq:square_new} \\
    & \sqrt{ \left[ \ln \frac{a}{r_0} - 1.3105 \right]^2 + 6.1064 \cdot 10^{-1} } \; \Bigg\}
    \nonumber 
\end{align}
where the expression $\ln \left( a / r_0 \right) - 1.3105$ is equal to the denominator of formula (I.5) (see Tables \ref{tab:overview} and \ref{tab:constants}).

In Fig. \ref{fig:eps_compare}, we compared the performance of several formulae from Table \ref{tab:overview} and the obtained Eq. (\ref{eq:square_new}) by calculating the relative error of the plasma frequency estimations compared to \textit{the exact values}.

For thin wires (see Fig. \ref{fig:eps_compare}(a)), up to the $r_0 / a$ value of $0.1$, Eq. (\ref{eq:square_new}) perfectly matches the results of Eq. (I.2) by P.A. Belov and outperforms all other formulae from Table \ref{tab:overview}. The largest error for Eq. (\ref{eq:square_new}) is $\sim 0.15\%$ at $r_0/a \sim 0.08$. For example, the direct solution of the transcendental equation (I.2) yields an accuracy of $\sim 0.5\%$ at $r_0 / a = 0.1$ (for $r_0/a < 0.1$, the error is lower).

In the range of thick wires ($r_0/a > 0.1$), the estimation (I.9) by A. Kumar et al. still performs best, providing $<2.5\%$ accuracy in plasma frequency estimation up to an $r_0 / a$ value of $0.3$, which corresponds to relatively large wire radii -- see Fig. \ref{fig:eps_compare}(b). For wires radii less than $0.1a$ this approximation exhibits an almost constant error of a fixed percentage $\sim 0.2$–$0.5\%$.

For thick wires the proposed Eq. (\ref{eq:square_new}) gives a more accurate estimation than the transcendental equation (I.2), with an accuracy of $\sim 2.5\%$ up to an $a/r_0$ value of $0.2$.

We also tested the performance of Eq.~(\ref{eq:rect_root}) for a rectangular wire medium ($a < b$) and compared it with numerical (\textit{exact}) results.

Originally, Eq.~(I.5) from Table \ref{tab:overview} was derived (see Eq.~(37) in \cite{belov2002dispersion}) for a rectangular lattice, and in our notation it can be written as:
\begin{equation}
    k_p^2 = \frac{2\pi/ab}
    {\ln{\frac{\sqrt{ab}}{r_0}} + F_1(b/a)}.
    \label{eq:belov_rect}
\end{equation}
We include this expression in the comparison below.

\begin{figure}[t]
    \begin{minipage}{1.0\linewidth}
		\center{ \includegraphics[width=1.00\textwidth]{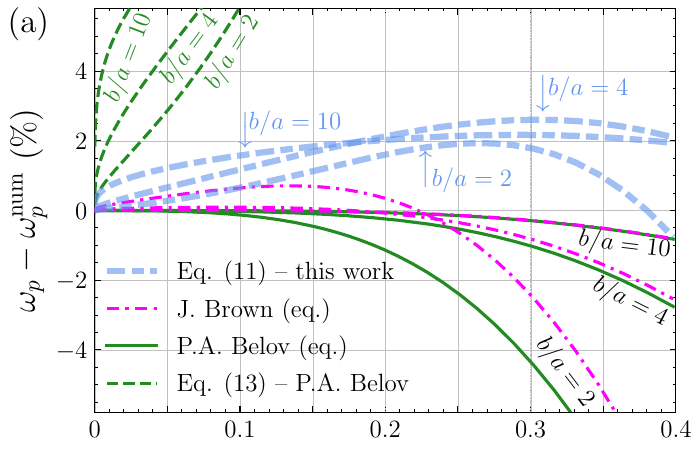}
		}
	\end{minipage}
    \hfill
    \begin{minipage}{1.0\linewidth}
		\center{ \includegraphics[width=1.00\textwidth]{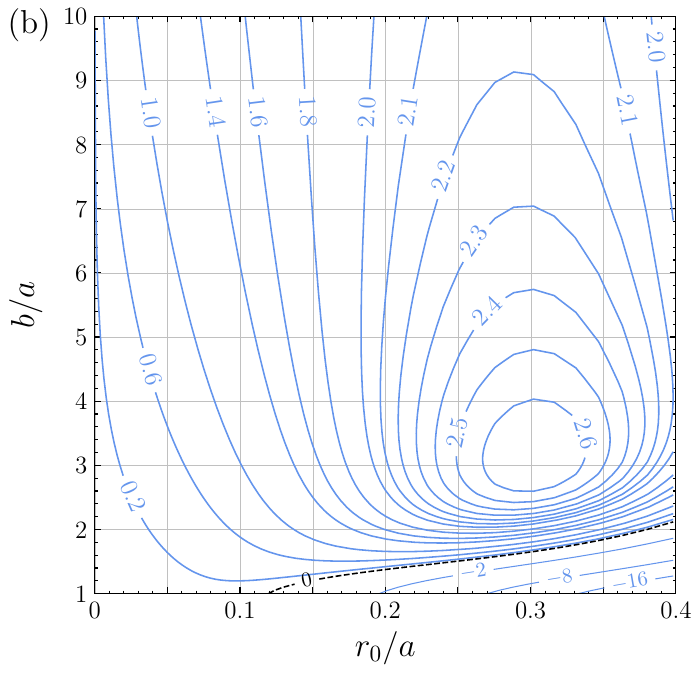}
		}
	\end{minipage}
	\caption{
        (a) Relative error (in percent) with respect to the \textit{exact} (numerical) results for the various formulae used to estimate the plasma frequency of a wire medium with \textbf{a rectangular lattice}.
        (b) Relative error (in percent) isocontours for the new formula (\ref{eq:rect_root}) plotted in $b/a$ versus $r_0/a$ parametric space.
	} \label{fig:rect} 
\end{figure}

Figure~\ref{fig:rect}(a) shows the comparison of Eq.~(\ref{eq:rect_root}), Eqs.~(I.1–2), and Eq.~(\ref{eq:belov_rect}) for tree different values of the period ratio $b/a>1$ ($b/a=2$, $4$ and $10$). Similar to Fig.~\ref{fig:omega_comsol}(c–d) and Fig.~\ref{fig:eps_compare}, the plot in Fig.~\ref{fig:rect}(a) illustrates how the relative error of each approximation depends on the wire radius $r_0$. 
The results indicate that the estimates obtained with the proposed formula (\ref{eq:rect_root}) are more accurate than those from Eq.~(\ref{eq:belov_rect}) for all $b/a$ ratios.
Although the transcendental equations (I.1–2) provide the most accurate estimates of the plasma frequency as $b/a$ increases, they are not straightforward to use and must be applied with care (for example, a simple permutation $a \leftrightarrow b$ produces a different result).


We also evaluated the relative error of our formula (\ref{eq:rect_root}) against the \textit{exact} numerical results over the parameter range $1 \leq b/a \leq 10$ and varying wire radii $r_0$. 
The isocontours in Fig.~\ref{fig:rect}(b) illustrate how the relative error changes across this parameter space. 
The black dashed line indicates the parameter values for which the estimate coincides with the exact result (i.e., the relative error is zero).
The proposed formula yields less than $2.7\%$ error overall in the range $2 \leq b/a \leq 10$ and $r_0/a \leq 0.4$ (and less than $1.5\%$ for $r_0/a \leq 0.1$).

In conclusion, we proposed a new formula (\ref{eq:square_new}) for estimating the plasma frequency of a simple wire medium with rectangular unit cell, derived using the line current approximation \cite{belov2002dispersion}. This formula is practically usable and does not require solving of any transcendental equation and only require computing two series which converge fast.

The performance of the proposed formula was evaluated for rectangular lattices over a wide range of wire radii and the ratio of lattice periods $a$ and $b$ by comparison with numerically obtained eigenfrequencies and with existing estimations from a scholarly literature. 

We demonstrated that the new formula provides the most accurate results \textbf{for a square lattice} when wires radii $r_0$ is more than $10$ times smaller than the lattice period, yielding a relative error of less than $0.16\%$.

\textbf{For a rectangular lattice} with the periods ratio varying from $1$ to $10$ and wire radius changing up to $0.4$ of the smallest period, the relative error of the estimation provided by the proposed formula compared to the numerically obtained plasma frequencies does not exceed $2.7\%$.







\nocite{*}

\bibliography{apssamp}

\end{document}